# Finding an Accurate Early Forecasting Model from Small Dataset: A Case of 2019-nCoV Novel Coronavirus Outbreak

Simon James Fong[1,2]*, Gloria Li[2], Nilanjan Dey[3]*, Rubén González Crespo[4], Enrique Herrera-Viedma[5]

[1] Department of Computer and Information Science, University of Macau, Macau SAR (China)
[2] DACC Laboratory, Zhuhai Institutes of Advanced Technology of the Chinese Academy of Sciences (China)
[3] Department of Information Technology, Techno India College of Technology (India)
[4] Universidad Internacional de La Rioja, Logroño (Spain)
[5] University of Granada (Spain)

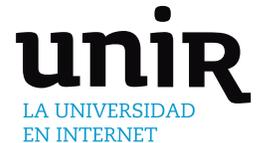

## Abstract

Epidemic is a rapid and wide spread of infectious disease threatening many lives and economy damages. It is important to fore-tell the epidemic lifetime so to decide on timely and remedic actions. These measures include closing borders, schools, suspending community services and commuters. Resuming such curfews depends on the momentum of the outbreak and its rate of decay. Being able to accurately forecast the fate of an epidemic is an extremely important but difficult task. Due to limited knowledge of the novel disease, the high uncertainty involved and the complex societal-political factors that influence the widespread of the new virus, any forecast is anything but reliable. Another factor is the insufficient amount of available data. Data samples are often scarce when an epidemic just started. With only few training samples on hand, finding a forecasting model which offers forecast at the best efforts is a big challenge in machine learning. In the past, three popular methods have been proposed, they include 1) augmenting the existing little data, 2) using a panel selection to pick the best forecasting model from several models, and 3) fine-tuning the parameters of an individual forecasting model for the highest possible accuracy. In this paper, a methodology that embraces these three virtues of data mining from a small dataset is proposed. An experiment that is based on the recent coronavirus outbreak originated from Wuhan is conducted by applying this methodology. It is shown that an optimized forecasting model that is constructed from a new algorithm, namely polynomial neural network with corrective feedback (PNN+cf) is able to make a forecast that has relatively the lowest prediction error. The results showcase that the newly proposed methodology and PNN+cf are useful in generating acceptable forecast upon the critical time of disease outbreak when the samples are far from abundant.

## Keywords

Forecasting, Machine Learning, Method, Prediction, Data Mining, Epidemic.

## I. Introduction

Since December 2019, the first case of human infection by a novel (unknown and new) virus which was formerly and now known as Wuhan virus and Coronavirus coded as 2019-nCov respectively, was reported [1]. As it was speculated that 2019-nCov originated from a single wild-animal which was traded at a busy marketplace [2], the toll of infested people sharply arose in the city Wuhan, spreading to other Chinese cities and eventually becoming a global epidemic within merely a month. 2019-nCoV is found to be highly contagious via bodily droplets in air, the virus can live up to hours to two days when left over a contacted surface. Human-to-human transition has recently been detected, and the toll of the infested is rising up almost exponentially in this early stage. A global health emergency warning was issued by the World Health Organization on 30 January 2020 [3], designating that 2019-nCoV is of an urgent global concern. At the early stage, the morbidity mortality rates resulted from the infection of 2019-nCoV are uncertain [4] especially for young children and the senior aged groups. In order to control the wide and fast spread of the virus, authorities took pre-emptive measures to cordon off infested cities. Such measures include closing borders, suspending community services and schools, minimizing both domestic and international travels etc., until further notice. The purpose is to limit the chances of physical contacts among people, so to thwart the transmission of the novel virus.

The curfew imposed for China and other countries will cost huge economical loss increasingly every day. As this is a novel virus, its severity is unforeseen albeit its contagiousness is very strong and the incubation period is relatively longer than other virus, it is difficult for anybody to decide the optimal time of lifting the ban. Too soon should the ban be lifted; the epidemic may not have totally subsided; prolonged extension in constraining the societies will lead to heavier

* Corresponding author.

E-mail addresses: ccfong@um.edu.mo (S. J. Fong), nilanjan.dey@tict.edu.in (N. Dey).

economical loss. Timing is very uncertain during this initial stage provided that the virus is novel, and we all have little knowledge about its characteristics. Authorities want to know when this epidemic may end, and whether it is continuing to get worse.

Therefore, forecasting is extremely important even for the slightest clues for multi-attributes consideration over other societal and public health factors. In this case, just a forecast from any single model is not enough, the most reliable forecast is needed often from selection of several candidate models. In data mining this is a challenging computational problem [5] [6] - how to build a most accurate forecasting model with only a handful of training data available in the early stage? Three popular approaches of building prediction model from a small dataset are reported in the literature [7] [8]. One way is to expand the training dataset by augmenting additional data onto the available data [9] [10]. The second way [11] is to use an ensemble for collectively generating forecast results, and the forecast result from one of the several algorithms that has the lowest error is selected to use. The results from the rest of the candidates are discarded. The third approach [12] is to stay focus on a single prediction algorithm which often comes with several parameters to be adjusted. The accuracy level of the resultant model tends to be very sensitive to parameter setting. The default parameters values for such algorithm often do not provide the maximum performance, fine-tuning at the parameters values is required to improve the accuracy level.

In view of the gravity that associated with decisions to be made about the epidemic, a new data mining methodology should be considered as an alternative to the existing ones on inferring the best model from few training data. In this paper, the newly proposed methodology embraces the merits of the three methods with the following features: 1. multiple candidate forecasting algorithms are put together for group prediction, the one that has the lowest error is selected; 2. each of the forecasting model is tuned with the most suitable parameter values; and 3. relevant information about the prediction target are added in multiple regression type of forecasting models, as one of the group selection candidates.

The reminder of the paper is structured as follow. Section two describes the new group forecasting model with early stage small dataset samples. The experiment is presented in Section three, with discussion of results follows on Section four. Section five concludes the paper as well as sharing with future works for follow-up research on this critical topic.

## II. OUR PROPOSED METHODOLOGY

In the early stage of decision making for fast developing epidemic, little data is available, our forecasting model needs to deal with high uncertainty. It is assumed that the virus is novel and human expert judgement will be consulted after a technical forecast is made, similar to Delphi decision making. A methodology for data mining using small dataset is needed in consideration of three main objectives - the adopted forecasting model must be most competitive compared to its peers (with the lowest error), the wining model itself is optimized to its maximum performance, and the wining model has the flexibility of including other relevant time-series for multiple regression.

Our proposed methodology is designed with respect to offering the highest possible level of prediction accuracy under the constraints of low data availability and knowledge. The design focus on doing group forecasting with a collection of optimized forecasting models, some of which are able to take multiple data sources as inputs. The methodology is therefore named as Group of Optimized and Multi-source Selection, abbreviated as GROOMS.

### A. Group of Optimized and Multi-source Selection

As a methodology, GROOMS has three main processes. Firstly, the full dataset, however small it may be, is passed to a collection of candidate forecasting models. Secondly each candidate model will be tuned to its best performance by optimizing its parameters values or inclusion of data sources. In some special case of multiple regression, more than one data sources which are related to the prediction target are fed into the inputs of a multi-variable prediction model for reasoning. Neural network is one typical example. Depending on the characteristics of the forecasting model, which are roughly classified into three: non-parametric (e.g. linear regression, simple statistical inference) that requires no setting of model parameter; parametric model (e.g. choice of activation function, learning rate, for neural network, entropy, pruning option and splitting criteria for decision tree) in which the output performance is sensitive to the choice of the parameters values; and dual models whose performance is very sensitive to both choices of input sources and the model parameters setting. The steps and processes of GROOMS are shown in Fig. 1.

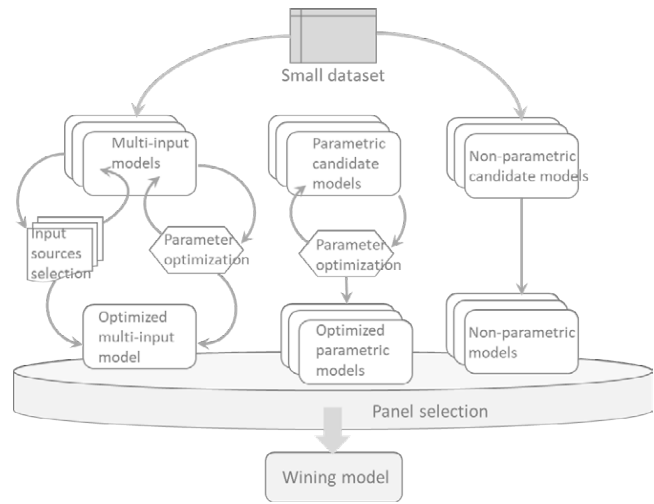

Fig. 1. GROOMS with top-down optimization processes.

In GROOMS, the available small dataset is passed from top to down through optimization processes, preparing the candidate models for panel section. Taking the curve fitting error by a forecasting model as the prime performance indicator, the panel selection picks a winning model that has the lowest error. The prediction result from the winning model is hence considered as the best effort prediction out of what technically is available using simple desktop forecasting technology [13].

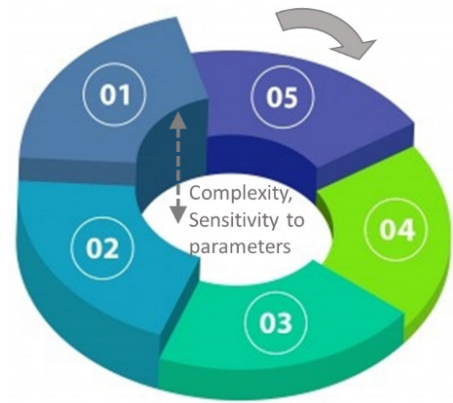

Fig. 2. GROOMS in recommended working order and complexity.

In general, there are five groups of data analytics pertaining to constructing a forecasting model. As shown in Fig. 2, five groups embrace different types of algorithms and they have different complexity in the model designs [14].

**Group 01**: Forecasting using complex machine learning models with multiple regressions.

This is usually the neural network type of models which have highly configurable structures and parameters, they usually work on multi-variate input data. When the number of attributes or sources of the input data is high, the data is said to have high dimensionality and feature selection [15] may need to be applied. In the context of early stage epidemic forecasting where high dimensional data may not be so available, feature selection is a matter of manually selecting a combination of appropriate time-series to feed into the learning model. This group of candidate models requires dual tuning of model parameters together with the selection of a subset of time-series for minimizing the prediction errors.

**Group 02**: Forecasting using complex machine learning models.

This is referring to forecasting models which use relatively complex machine learning models as base learners. The input data are ordered with timestamps, the base learners attempt to learn the patterns and output predictions or forecasts by regression. Some example is support vector machine or neural network which requires iterative hyperplanes/weights updates for maximizing the results. Normally this type of models consumes considerable amount of memory and runtime.

**Group 03**: Forecasting using light-weight machine learning models

Similar to the one about, but it uses simple machine learning models as base learners. They run faster and generate results almost as quickly as traditional time-series forecasting. However, underfitting and overfitting can easily occur. Parameter tuning should be done to assure a reliable model.

**Group 04**: Simple data analytics

Mostly are descriptive statistics, e.g. average, min, max, the increments since yesterday or some days ago, rate of increases of suspected cases in comparison to cured cases, etc.

**Group 05**: Econometrics type of time-series forecasting

Traditional econometrics such as auto regression, exponential regression, moving average, ARIMA and its variants etc. These methods usually work over univariate time-series with none or little parameter. For example, window size for moving average, and ratios between autoregression and moving average for ARIMA, etc.

The five groups of methods have various complexity therefore run-time and modeling time which may require manually fine-tuning which takes time. In view of urgency when a quick but most reliable possible forecast is required, it is suggested that in GROOMS the groups of methods should be attempted in priorities in this sequence, as shown in Fig. 2. That is, traditional forecasting from group 5 should be first applied, probably at the same time with group 4 statistical analysis. When resources and spare time are available, methods from groups 3, 2 and 1 could be attempted in order for pursuing a lowest cost forecasting model under time constraints.

*B. Polynomial Neural Network with Corrective Feedback (PNN+cf)*

For group 1 - Forecasting using complex machine learning models with multiple regressions, wide choices of candidate models are available. They range from a simplest single perceptron learner to sophistical deep learning models that have many configuration options. In the literature there are state-of-the-arts that discussed using heuristic techniques for hyper-parameter optimization. It is generally known that the more parameters involved in a machine learning model, the more variables to the influences of the ultimate performance. Epidemic need quick decisions, and it could be a tedious task to try out every combination of parameter values for every model even using semi-automated approach. To this end, an alternative neural learning approach called Polynomial Neural Network is chosen in lieu of mainstream convolution neural network.

For this forecasting task of early stage epidemic, an evolving deep learning model is used. In an early stage, we neither know about in advance how long nor how fluctuating the time-series will be as time goes by. Likewise, it is not known and therefore should not permanently fix or limit about the structure of a neural network in terms of how many hidden layers and number of neurons in each layer that the network should scale up to. Therefore, an evolutionary type of neural network is desirable for such situation.

In this study, a type of evolutionary neural network called polynomial neural network (PNN) which is based on the classic principle of Group Method Data Handling (GMDH) is used. PNN expands the neural network complexity iteratively until no more performance improvement by further addition of neuron to the network structure is observed.

The original concept which was invented in 1968 by Ivakhnenko [16], is known as one of the oldest prototypes of convolution neural network. PNN is mainly powered by idea of complicating a polynomial equation which grows a neural network with scaling up the powers of the polynomial coefficients until it can model the time series as fit as possible. As a result, the neural network would remain at marginal model complexity while its predictive ability is at its highest. PNN grows the network structure by self-tuning the model's coefficients (parameters) automatically through an iterative data sampling and controlled network expansion process. The network size is therefore increasing one iteration at a time. This is a typical non-linear optimization process between the inputs and the output of the network. In its simplest form, the optimization argument by increasingly complicating the current model is as follows:

$$\bar{\beta} = \arg\min_{\beta \subseteq B} Output\_error(\beta) \qquad (1)$$

where $Output\_error(\beta)$ is a criterion of output error of the model $\beta$, which will be iteratively tried from a full space of candidate beta models $B$. The candidate beta models are the components that build up a final PNN which piece themselves into a nonlinear multi-parametric equation. The equation $\beta = f(x_1, \ldots, x_n)$ maps the non-linear relation among a range of sequential variables, $x_{t=1}, \ldots, x_{t=n-1}, x_n$ which is comprised as an input vector $\bar{x}$, and the predicted outcome is y from the final model $\bar{\beta}$.

Hence a PNN is simply a polynomial equation whose coefficients and their powers are modelled as weights and neurons. To compute the values of the coefficients for solving the polynomial equation, a polynomial reference function which is Equation 2 is used that represents the different candidate models that grow in PNN optimization process as follows:

$$y = \varepsilon_0 + \sum_{t=1} \varepsilon_{t=1} x_{t=1} + \sum_{t=1}\sum_{t=2} \varepsilon_{t=1,t=2} x_{t=1} x_{t=2} + \sum_{t=1}\sum_{t=2}\sum_{t=3} \varepsilon_{t=1,t=2,t=3} x_{t=1} x_{t=2} x_{t=3} + \cdots \qquad (2)$$

Equation 2 shows a classical Kolmogorov-Gabor polynomial which resembles a functional series for PNN; it is able to represent any function in a general form $y=f(\bar{x})$ with its coefficient vector $\bar{\varepsilon}$. $\bar{\varepsilon}$ is the solution which PNN tries to obtain by solving the regression polynomial. As time goes that is modelled by iteration, the values of variables $\bar{x}$ feed to PNN, it computes the coefficients at the same time trying to incrementally increase the complexity. This is controlled by monitoring the error level. When there is no significant incremental gain in performance, y is then taken as the forecast result.

There are two optional forms of PNN, one being Combinatorial PNN (Combi) [17] and the other one is Multilayer Iterative PNN-type Neural Network (MIA). Combi simulates the polynomial formulae by a simple combinatorial function that rolls up the power iteratively and extends the polynomial equation with higher power. MIA [18] is modified from a standard feed-forward neural network which is constructed initially from bi-input neurons. The neurons and layers expand by some neuron selection criteria in a way similar to the design reported in [19].

PNN has been applied for predictions of massive earthquake disasters [20], energy consumption [21] and gas furnace behaviours [22] which all are quite erratic and hard to predict in nature. To further improve the capability of PNN, a new version called PNN with corrective feedback (PNN+cf) is proposed.

PNN+cf allows feeding in extra input variables during the formation of polynomials. However, it should be carried out carefully because redundant variables can lead to overfitting, thereby degrading accuracy, and incurring long computational time cost. Only statistically significant or so-called relevant extra information should be added for accuracy enhancement. In this study two extra information, lagged data and training errors from past iterations of model training which are known as corrective feedbacks are added. The feedbacks are 1-step lag which appears like a shadow of the time series and residuals. The motivation for adopting residuals which are the deviations between the predicted and actual values, is the compensation of bias towards the nonlinearity that exists in the time series.

An illustration of the fusion of the two extra inputs to a neural network is shown in Fig. 3. The inputs from the original time series, values from the lagged data series, and residual information are augmented together. A rolling window of certain size temporarily feeds a part of the training prior to the neural network, and it predicts an outcome one step ahead. At each prediction, the residuals from previous training cycles are fed back to the neural network. By this way, the regression process embraces both the fresh input data while adjusting their weights that are biased away from the residuals. A better fitting curve is resulted from the polynomial equation for regressing a low-error output.

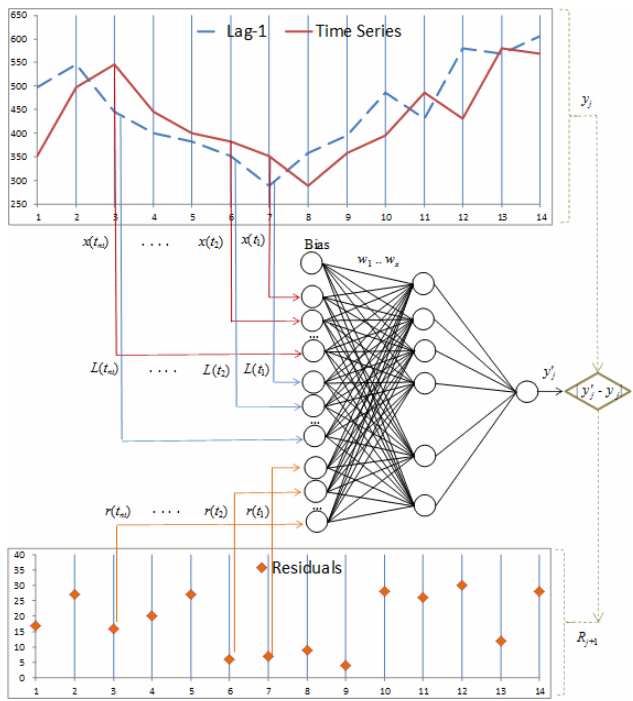

Fig. 3. Additional inputs being fed into PNN-cf network.

PNN+cf is applied here for demonstrating a scenario of using a neural network which can adaptively and dynamically adjust its structure via an optimization process and input sources selection.

### III. Experiment

An experiment is carried out to verify the efficacy of the proposed GROOMS, particularly for early stage forecasting of the 2019-nCov epidemic. Empirical data is used, which is extracted from the archive of Chinese health authorities. The data is updated daily since 21 Jan 2020, of the daily increment of the number of infested people in China. The trends are put in categories of which makes up a full cycle of a case after the infection: Confirmed, Cured, Died, Suspected, Critical. A patient is deemed as suspected when s/he started to show symptoms of virus infection; through some medical diagnosis afterward, s/he will be confirmed or clear. The patient then may fall into critical condition which s/he would be either cured or died. These figures show daily trend as the epidemic develops. Till the time of writing, the epidemic trend is still on the rise, though for experiment only the data between 21 Jan – 3 Feb 2020 are used. A time series of 14 instances about the Suspected cases is to be run through GROOMS for 6-days ahead forecast. At any point of time in the progress of the epidemic, authorities are concerned about forecast of future days. The fourteen instances represent a challenging scenario of data mining over small data.

The experimentation platform is a i7-CPU running 64-bits OS of MS Windows 10, with installations of several machine learning benchmarking software – Weka, SPSS, NunXL, XLMiner, etc. from Waikato University, NA, Microsoft Inc, IBM Inc, Spider Software Inc and Solver Inc. respectively. In most cases, default parameters values are used as they are to start with the initial run.

Besides the fact that the curve of Suspected is non-stationary with a rising trend apparently, the technical decomposition is technically challenging – it has a sharp hump and trough near the end of the first week, followed by another smaller reverse trough and hump towards the end of the curve. Intuitively there are no elements of seasonality, cycle, and white noise. Although small the available data amount is, it has partially some but incomplete characteristics of logarithmic curve and repetitive zigzags. From humanity point of view, daily increase of suspected cases might be an important indictor of a wide-spreading epidemic (see Fig. 4).

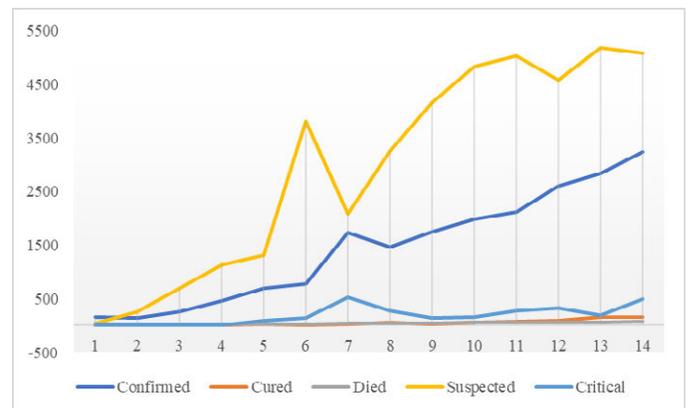

Fig. 4. Daily increase of Wuhan coronavirus infection cases.

Three representative groups of forecasting algorithms are put into testing of GROOMS. They are classical time series forecasting algorithms, machine learning models and polynomial neural networks for time series forecasting respectively. The rational is to observe and investigate how matching the curve fitting produced by these algorithms to the limited epidemic data at the early stage. This is to simulate a situation when the outbreak just occurred, how these forecasting algorithms perform at their best under crisis. Root-mean-square-error (RMSE) which is a standard benchmarking performance measure in time series forecasting is used as a criterion in panel selection.

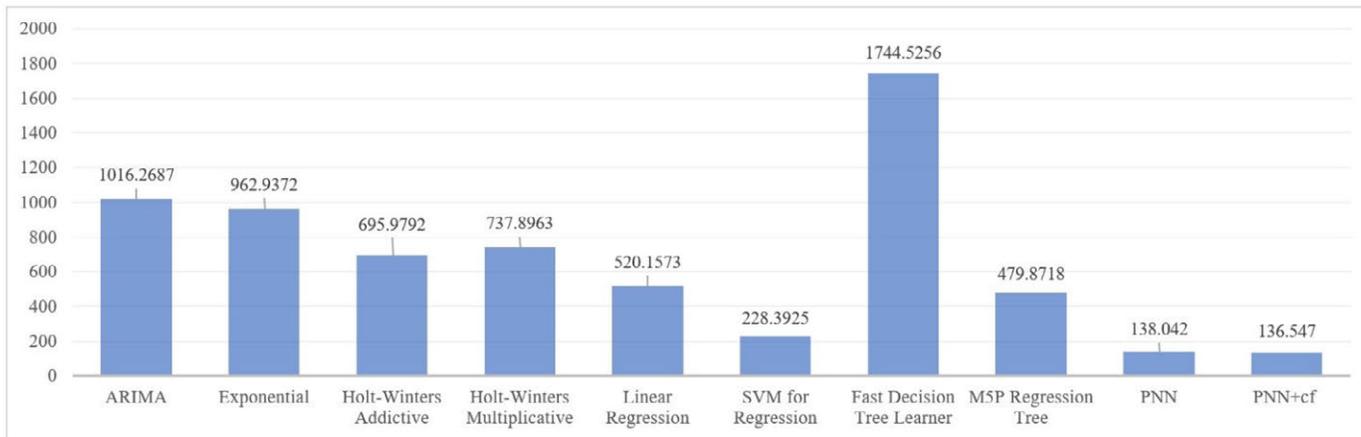

Fig. 5. RMSE comparison between three groups of forecasting methods – traditional (left), machine learning (middle) and PNN (right).

The same small dataset is loaded into various above-mentioned algorithms without any pre-processing for competition. If automatic tuning is available for the algorithm, it is activated, or else default parameter values would be assumed. At the end of their final runs, RMSE was recorded down for comparison. They are shown in bar charts as in Fig. 5.

The result bars are positioned in such a way that the classical algorithms are the left, the machine learning algorithms are in the middle, and the neural network-based algorithms are at the right. It can be seen that in general the classical time series forecasting algorithms failed to perform well. There are marginally large gaps mainly between the fitting curve and the actual data at the hump near the initial stage. Please see Fig. 6 a-d.

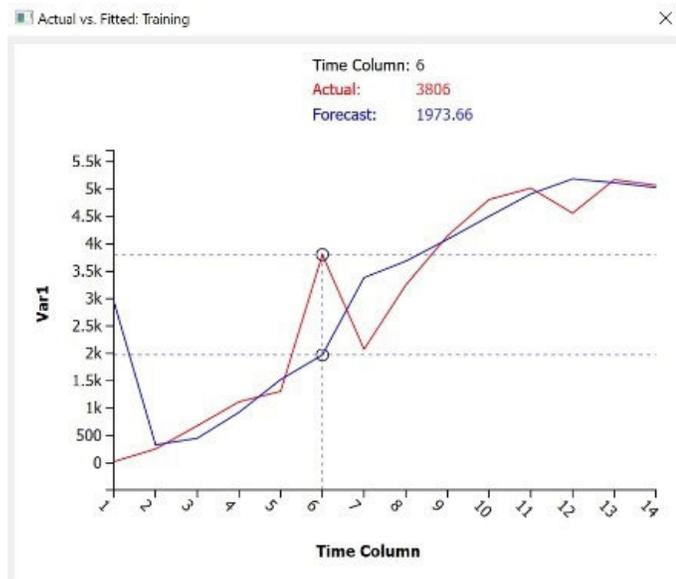

a) ARIMA

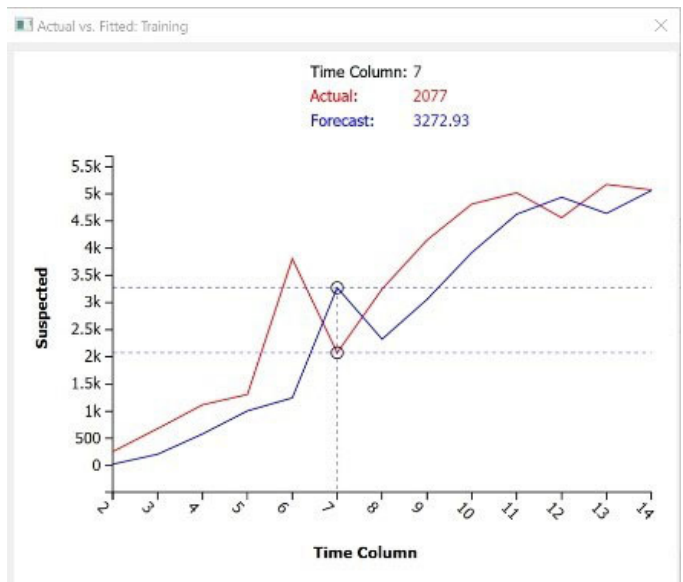

b) Exponential

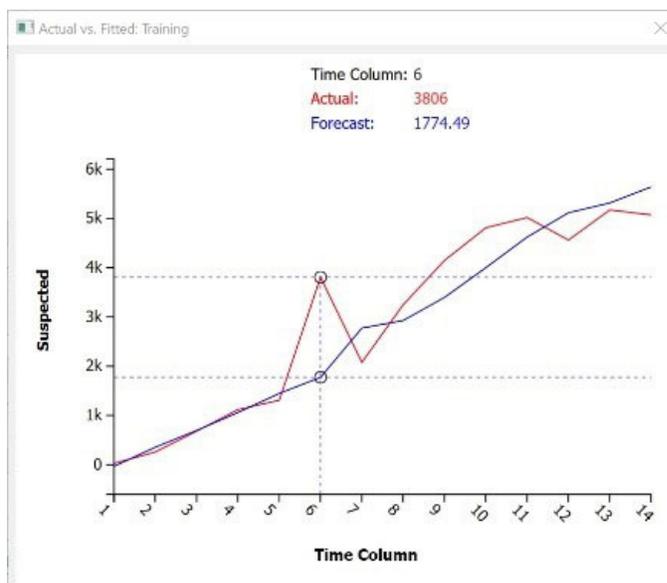

c) Holt-Winters Addictive

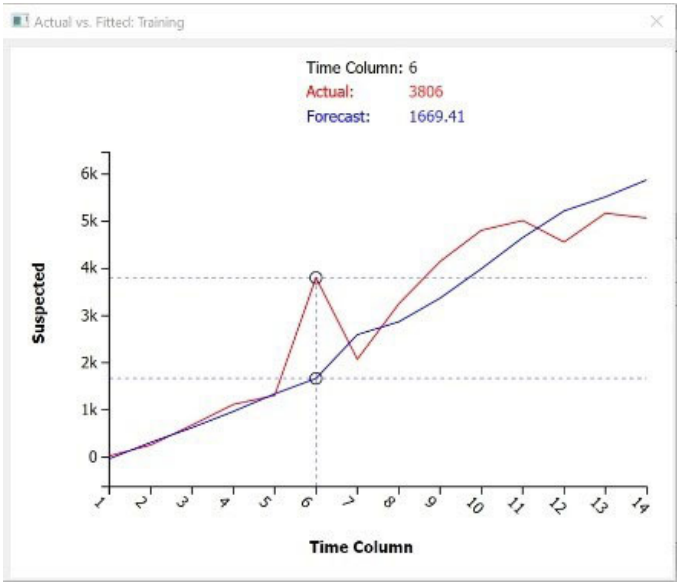

d) Holt-Winters Multiplicative

Fig. 6. Deviation between predicted and actual values.

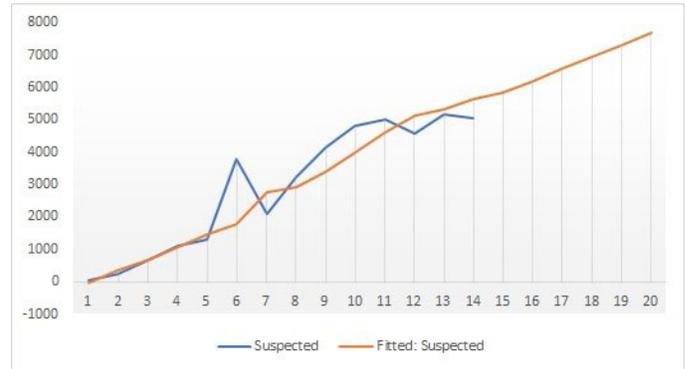

c) Holt-Winters Addictive

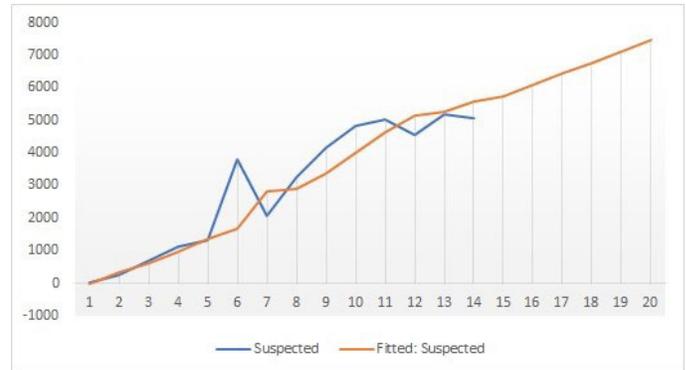

d) Holt-Winters Multiplicative

Fig. 7. Forecasts generated by traditional time series forecasting algorithms.

Such drastic changes at the curve hampered the capability of algorithms that are based on curve-fitting. Consequently, by referring to the rates of RMSE, machine learning models appear to be more promising than those classical algorithms for this particular case. An exception is Fast Decision Tree Learner that scored RMSE 1744.526. On the other hand, neural network based PNN's are found to perform the best in this panel selection at the lowest RMSE 136.547 observed. The forecasting results by traditional time series forecasting methods - ARIMA, Exponential, Holt-Winters Addictive and Hot-Winters Multiplicative are shown in Fig. 7 a-d. The forecasts by machine learning models are shown in Fig. 8 a-d.

It is observed that the four forecasting results from Fig. 7 a-d that they basically give very different trends. Holt-Winters' up, ARIMA down, and Exponential flat. Just by judging the RMSE within this group of predictors alone, Holt-Winters has the lowest error. So, it is indicating a rising trend that with an almost constant gradient. However, at the panel selection of GROOMS, there are forecasting models by machine learning methods that have lower errors.

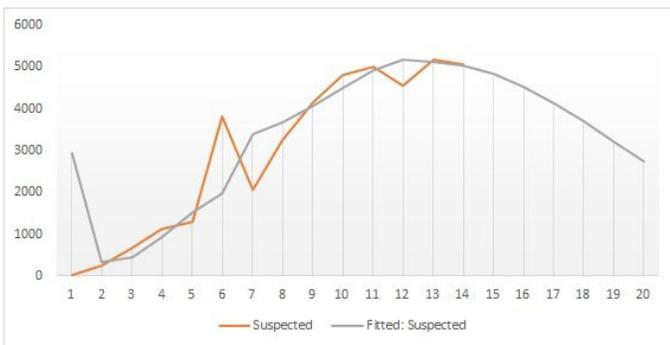

a) ARIMA

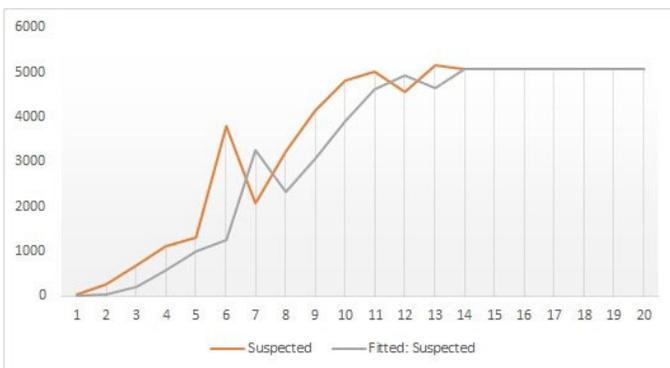

b) Exponential

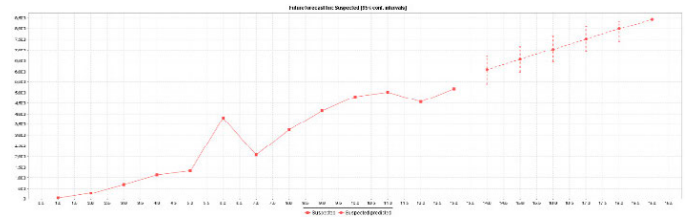

a) Linear Regression

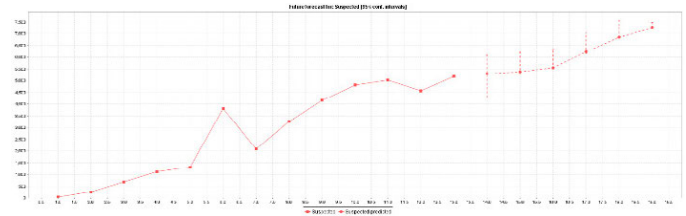

b) SVM for regression [23]

TABLE I. RMSE of Various Types of PNN coupled with Different Subsets of Input Data Sources

| Polynominal NN types | All | suspected + confirmed | suspected + confirmed + critical | suspected + confirmed + critical + cured | suspected + confirmed + critical + died | suspected + critical | suspected + cured | suspected + died |
|---|---|---|---|---|---|---|---|---|
| Combi | 200.044 | 190.599* | 190.599* | 237.261 | 214.097 | 193.520 | 276.885 | 290.079 |
| Combi-cf | 174.366 | 189.661 | 160.338* | 165.127 | 172.627 | 167.544 | 241.052 | 222.997 |
| MIA | 153.709 | 155.684 | 138.042* | 141.557 | 156.275 | 158.156 | 183.904 | 186.212 |
| MIA-cf | 151.789 | 154.128 | 136.547* | 140.780 | 157.496 | 162.175 | 189.954 | 186.212 |

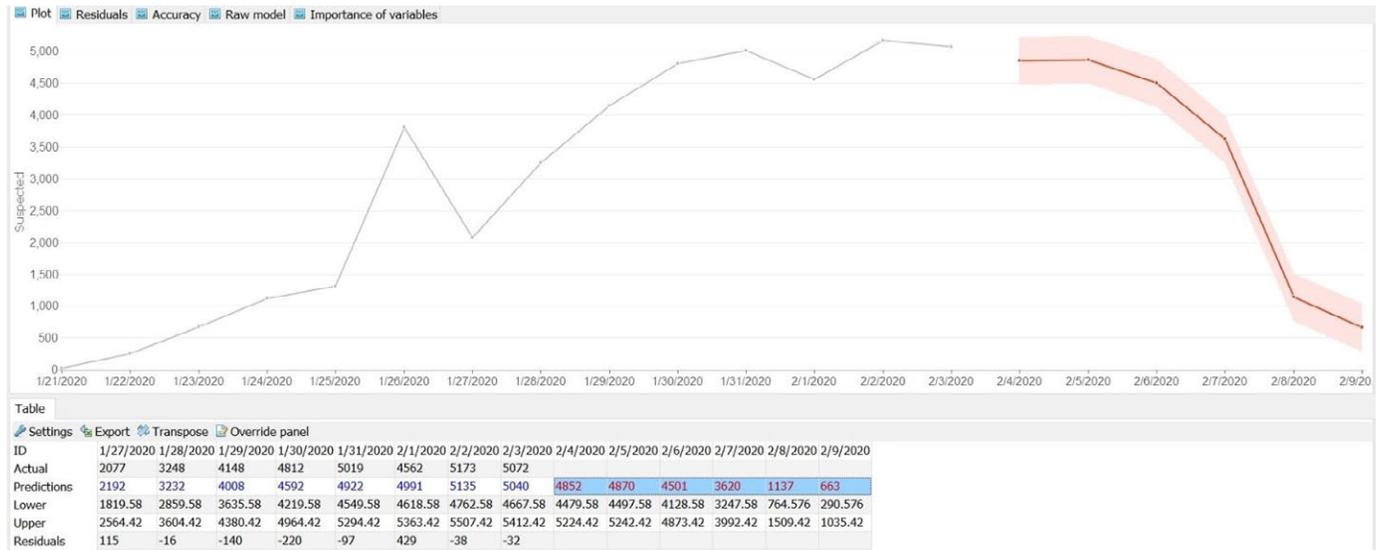

Fig. 9. Forecasts generated by PNN+cf algorithms.

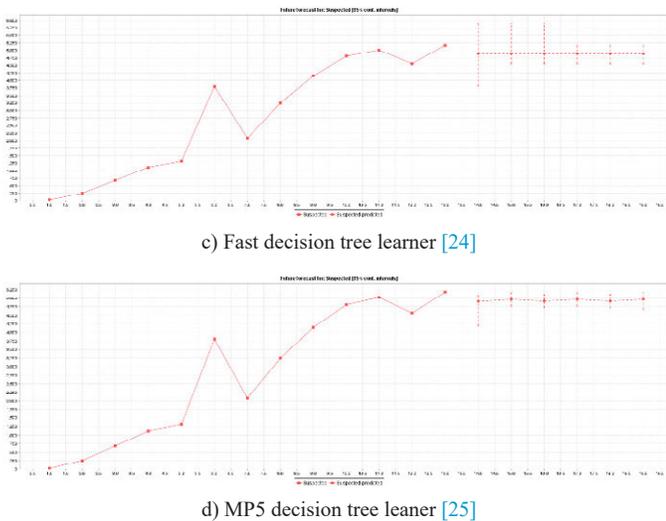

c) Fast decision tree learner [24]

d) MP5 decision tree leaner [25]

Fig. 8. Forecasts generated by machine learning algorithms.

The forecasts by this group which are shown in Fig. 8 appear like straight lines except SVM for Regression. This suggest that SVM which is a strong hyperplane classifier capable of recognizing very non-linear pattern (in this case, the epidemic trend is indeed curvy and non-smooth), it is able to produce a forecast with certain blends and turns, based on the past history which is anything but smooth.

The winning model from panel selection is PNN+cf which evolved its structure to its most optimal. It belongs to Group 1 according to GROOMS, which allows inclusion and selection of multiple data sources other than the one that is being forecasted. A separate experiment is therefore conducted that tried several combinations of data sources; from there we observe which combo has the lowest error rates. Please refer to the results in Table I. By default, when all data sources are used together in the multiple regression, mediocre performance are obtained. It was found that, for both types of Combi and MIA architectures of PNN, this specific combination of data sources - suspected + confirmed + critical, when used together, superior performances are obtained. Intuitively, this is a natural sequence that resembles the lifecycle of an infested patient. An infection case that contributed to the national health records, started with somebody who displayed certain symptoms, hence a suspected case. After some diagnosis and test, the disease can be confirmed as coronavirus infection or something else. The patient received medical treatment; he may then turn critical that probably lead to fatality or cured. So, in order to forecast future suspected cases, the closely related status in sequence, that are confirmed and critical, could help. These three statuses form some kind of causality. They exist as a prerequisite for one to another to occur with high certainty. However, there exists a large uncertainty beyond the condition of critical at the moment, given that the virus is novel, its severity is not well-known, and the amount of available data is scarce. By the predicting powers of neural network, this uncertainty is reflected by the relatively higher RMSE on the right side of Table I.

Since the winning model PNN+cf offers the lowest RMSE among all, its forecasting result is shown in Fig. 9. If the data does not change (theoretically), the trend is forecast to dip down and the epidemic subsides. Of course, it is only a forecast as accurate as up to the current point of time, that is up to 4 Feb 2020. In this dynamic situation, the latest figures keep being updated daily, so does the forecast. Whenever new data comes in, it is highly likely that it will change the course of prediction, and a totally new outcome will be obtained which may be very different from the previous one. The future would unfold itself when time progresses.

## IV. Conclusion

Forecasting the ending time of an epidemic is always a noble but difficult task. From the perspective of data mining, the quest of finding the most accurate machine learning model while limited with very few data on hand to start about has attracted research attentions. In literature there are many bootstrapping and data augmentation methods for improving the performance of machine learning models. In this paper, a methodology called GROOMS is proposed which ensembles a collection of five types of forecasting methods, ranging from classical time-series forecasting, to self-evolving polynomial neural networks (PNN), to work together under a panel selector. In particular, our experiment has shown that PNN is superior in yielding a forecasting model with relatively lowest error. PNN is known to possess advantages in forming up a just fit model by incrementally growing its internal structure from simple to optimal. Therefore, over- or under-fitting is less likely to happen. In this paper, PNN is extended to PNN+cf with corrective feedback in the optimization process. At the best efforts in optimizing the parameters, input data sources for multiple regression using GROOMS, it is shown the possibility and feasibility in finding the most accurate forecasting model based on the limited availability of data. Although the experiment demonstrated the possibility of picking a suitable forecasting model, forecasting is a dynamic task itself whose predicted result is very sensitive to the parameters used, the choice of model and the training data. Any addition or modification on the training data upon the arrival of new data, is likely to influence the course of the predicted result (as seen in Fig. 7 – the uptrend vs downtrend).

In future, the underlying cause of the differences or deviation of forecasting results by different algorithms worth in-depth investigation. From data mining viewpoint, it is vital to test more other popular algorithms for panel selection. Better still, it worth the efforts in probing in the underlying designs of some prominent algorithms and understanding why or why not it incur a very low or very high errors. With this insight discovered, one may formulate even new breed of algorithms from inside out, which handles well small data yet giving acceptable quality forecast. The methodology should be extended too. So far GROOMS was constructed in mind that the accuracy/error level is the sole criterion. In reality, pondering whether an epidemic is going to end or escalate is a complex and wide multi-factors decision. Based solely on historical data is only of one aspect of contributions to the decision, though it is still important. A more thorough methodology would be useful for connecting the technical forecast to other non-technical decision-making processes. So, they can complement each other for making a multi-facet and reliable decision.

## References

[1] "WHO | Novel Coronavirus – China". WHO. Archived from the original on 23 January 2020. Retrieved 1 February 2020.

[2] Cohen, Jon (January 2020). "Wuhan seafood market may not be source of novel virus spreading globally". Science. doi:10.1126/science.abb0611. ISSN 0036-8075.

[3] "Statement on the second meeting of the International Health Regulations (2005) Emergency Committee regarding the outbreak of novel coronavirus (2019-nCoV)". World Health Organization (WHO). 30 January 2020. Archived from the original on 31 January 2020. Retrieved 30 January 2020.

[4] Sparrow, Annie. "How China's Coronavirus Is Spreading—and How to Stop It". Foreign Policy. Archived from the original on 31 January 2020. Retrieved 2 February 2020.

[5] Croda, R. M. C., D. E. G. Romero, and S. O. C. Morales, "Sales Prediction through Neural Networks for a Small Dataset", International Journal of Interactive Multimedia and Artificial Intelligence, vol. 5, no. 4, pp. 35-41, 03/2019.

[6] R. J. Hyndman, and A. V. Kostenko, "Minimum sample size requirements for seasonal forecasting models," Foresight, vol. 6, pp. 12-15, 2007.

[7] S. Ingrassia, and I. Morlini, "Neural network modeling for small datasets," Technometrics, vol.47, no. 3, pp. 297-311, 2005.

[8] A. Pasini, "Artificial neural networks for small dataset analysis." Journal of thoracic disease, vol. 7, no. 5, pp- 953, 2015.

[9] M. A. Lateh, A. K. Muda, Z. I. M. Yusof, N. A. Muda and M. S. Azmi, "Handling a Small Dataset Problem in Prediction Model by employ Artificial Data Generation Approach: A Review", Journal of Physics: Conference Series, Volume 892, Conf. Ser. 892 012016.

[10] R. Andonie, "Extreme data mining: Inference from small datasets," Int. J. Comput. Commun. Control, vol. 5, no. 3, pp. 280–291, 2010.

[11] T. Shaikhina, N. A.Khovanova, "Handling limited datasets with neural networks in medical applications: A small-data approach", Artificial Intelligence in Medicine, vol. 75, pp. 51-63, 2017.

[12] J. F. Slifker and S. S. Shapiro, "The Johnson system: selection and parameter estimation," Technometrics, vol. 22, no. 2, pp. 239–246, 1980

[13] R. Adhikari, and R. K. Agrawal, "An introductory study on time series modeling and forecasting," arXiv preprint arXiv:1302.6613, 2013.

[14] A. Singh, and G. C. Mishra, "Application of Box-Jenkins method and Artificial Neural Network procedure for time series forecasting of prices," Statistics in Transition new series, vol. 1, no. 16, pp. 83-96, 2015.

[15] Arun, V., M. Krishna, B. V. Arunkumar, S. K. Padma, and V. Shyam, "Exploratory Boosted Feature Selection and Neural Network Framework for Depression Classification", International Journal of Interactive Multimedia and Artificial Intelligence, vol. 5, no. 3, pp. 61-71, 12/2018

[16] A. G. Ivakhnenko (1970) "Heuristic Self-Organization in Problems of Engineering Cybernetics". Automatica Vol. 6, pp.207–219.

[17] A. G. Ivakhnenko, and A. A. Zholnarskiy, (1992) "Estimating the coefficients of polynomials in parametric GMDH algorithms by the improved instrumental variables method", Journal of Automation and Information Sciences c/c of Avtomatika, Vol. 25, no. 3, pp.25-32.

[18] S. K. Oh, W. Pedrycz, B. J. Park, "Polynomial neural networks architecture: analysis and design", Computers & Electrical Engineering, Vol. 29, No. 6, August 2003, pp. 703-725.

[19] A. Andoni, R. Panigrahy, G. Valiant, L. Zhang, "Learning Polynomials with Neural Networks", Proceedings of the 31 st International Conference on Machine Learning, Beijing, China, 2014. JMLR: W&CP volume 32.

[20] S. Fong, N. N. Zhou, R. K. Wong, X. S. Yang: Rare Events Forecasting Using a Residual-Feedback GMDH Neural Network. ICDM Workshops 2012: 464-473.

[21] N. Dey, S. Fong, W. Song, K. Cho (2018) Forecasting Energy Consumption from Smart Home Sensor Network by Deep Learning. In: Deshpande A. et al. (eds) Smart Trends in Information Technology and Computer Communications. SmartCom 2017. Communications in Computer and Information Science, vol 876. Springer.

[22] W. Waheeb, and R. Ghazali, Forecasting the Behavior of Gas Furnace Multivariate Time Series Using Ridge Polynomial Based Neural Network Models, International Journal of Interactive Multimedia and Artificial Intelligence, ISSN 1989-1660, vol. 5, no. 5, 06/2019, pp.126-133.

[23] S. K. Shevade, S. S. Keerthi, C. Bhattacharyya, K.R.K. Murthy: Improvements to the SMO Algorithm for SVM Regression. In: IEEE Transactions on Neural Networks, 1999.

[24] J. Su, H. Zhang, A fast decision tree learning algorithm, AAAI'06: Proceedings of the 21st national conference on Artificial intelligence – Vol. 1, July 2006, pp.500–505.

[25] Y. Wang, I. H. Witten: Induction of model trees for predicting continuous classes. In: Poster papers of the 9th European Conference on Machine Learning, 1997.

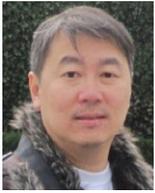
Simon James Fong

Simon Fong graduated from La Trobe University, Australia, with a 1st Class Honours BEng. Computer Systems degree and a PhD. Computer Science degree in 1993 and 1998 respectively. Simon is now working as an Associate Professor at the Computer and Information Science Department of the University of Macau, as an Adjunct Professor at Faculty of Informatics, Durban University of Technology, South Africa. He is a co-founder of the Data Analytics and Collaborative Computing Research Group in the Faculty of Science and Technology. Prior to his academic career, Simon took up various managerial and technical posts, such as systems engineer, IT consultant and e-commerce director in Australia and Asia. Dr. Fong has published over 450 international conference and peer-reviewed journal papers, mostly in the areas of data mining, data stream mining, big data analytics, meta-heuristics optimization algorithms, and their applications. He serves on the editorial boards of the Journal of Network and Computer Applications of Elsevier, IEEE IT Professional Magazine, and various special issues of SCIE-indexed journals. Simon is also an active researcher with leading positions such as Vice-chair of IEEE Computational Intelligence Society (CIS) Task Force on "Business Intelligence & Knowledge Management", TC Chair of IEEE ComSoc e-Health SIG and Vice-director of International Consortium for Optimization and Modelling in Science and Industry (iCOMSI).

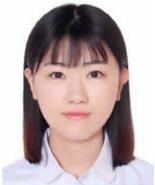
Gloria Li

Gloria Tengyue Li is currently a PhD student at the University of Macau. She is also the Head of Data Analytics and Collaborative Computing Laboratory, Zhuhai Institute of Advanced Technology, Chinese Academy of Science, Zhuhai, China. Ms Li is leading and managing the laboratory, in R&D as well as technological transfer and incubation. She is an entrepreneur with experiences in innovative I.T. contest, with her award-winning team in the Bank of China Million Dollar Cup competition. Her latest winning work includes the first unmanned supermarket in Macau enabled by the latest sensing technologies, face recognition and e-payment systems. She is also the founder of several Online2Offline dot.com companies in trading and retailing both online and offline. Ms Li is also an active researcher, manager and chief-knowledge-officer in DACC laboratory at the faculty of science and technology, University of Macau.

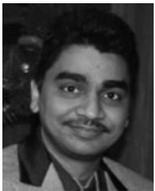
Nilanjan Dey

Nilanjan Dey, is an Assistant Professor in Department of Information Technology at Techno International New Town (Formerly known as Techno India College of Technology), Kolkata, India. He is a visiting fellow of the University of Reading, UK. He was an honorary Visiting Scientist at Global Biomedical Technologies Inc., CA, USA (2012-2015). He was awarded his PhD. from Jadavpur University in 2015. He is the Editor-in-Chief of International Journal of Ambient Computing and Intelligence, IGI Global. He is the Series Co-Editor of Springer Tracts in Nature-Inspired Computing, Springer Nature, Series Co-Editor of Advances in Ubiquitous Sensing Applications for Healthcare, Elsevier, Series Editor of Computational Intelligence in Engineering Problem Solving and Intelligent Signal processing and data analysis, CRC. He has authored/edited more than 50 books with Springer, Elsevier, Wiley, CRC Press and published more than 300 peer-reviewed research papers. His main research interests include Medical Imaging, Machine learning, Computer Aided Diagnosis, Data Mining etc. He is the Indian Ambassador of International Federation for Information Processing (IFIP) – Young ICT Group.

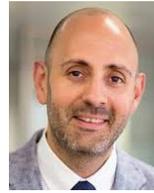
Rubén González Crespo

Dr. Rubén González Crespo has a PhD in Computer Science Engineering. Currently he is Vice Chancellor of Academic Affairs and Faculty from UNIR and Global Director of Engineering Schools from PROEDUCA Group. He is advisory board member for the Ministry of Education at Colombia and evaluator from the National Agency for Quality Evaluation and Accreditation of Spain (ANECA). He is member from different committees at ISO Organization. Finally, He has published more than 200 papers in indexed journals and congresses.

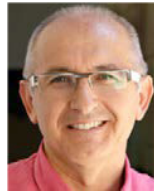
Enrique Herrera-Viedma

Enrique Herrera-Viedma is Professor in Computer Science and A.I in University of Granada and currently the new Vice-President for Research and Knowledge Transfer. His current research interests include group decision making, consensus models, linguistic modeling, and aggregation of information, information retrieval, bibliometric, digital libraries, web quality evaluation, recommender systems, and social media. In these topics he has published more than 250 papers in ISI journals and coordinated more than 22 research projects. Dr. Herrera-Viedma is Vice-President of Publications of the IEEE SMC Society and an Associate Editor of international journals such as the IEEE Trans. On Syst. Man, and Cyb.: Systems, Knowledge Based Systems, Soft Computing, Fuzzy Optimization and Decision Making, Applied Soft Computing, Journal of Intelligent and Fuzzy Systems, and Information Sciences.